\documentclass[slac_one]{revtex4}
\usepackage{graphicx}
\usepackage{fancyhdr}
\begin{document}

\title{Measurement of the Solar Neutrino Flux with an Array of Neutron
Detectors in the Sudbury Neutrino Observatory} 

%

\author{B. Jamieson (for the SNO collaboration)}
\affiliation{University of British Columbia, Vancouver, BC  V6T 1Z1, Canada}
\begin{abstract}

The Sudbury Neutrino Observatory has measured the $^8$B solar neutrino
flux using an array of $^{3}$He proportional counters.  Results
obtained using a Markov-Chain Monte-Carlo (MCMC) parameter estimation,
integrating over a standard extended likelihood, yield effective
neutrino fluxes of:
$\phi_{nc}=5.54^{+0.33}_{-0.31}$(stat)~$^{+0.36}_{-0.34}$(syst)$\times10^6$~cm$^{-2}$s$^{-1}$,
$\phi_{cc}=1.67^{+0.05}_{-0.04}$(stat)~$^{+0.07}_{-0.08}$(syst)$\times10^6$~cm$^{-2}$s$^{-1}$,
and
$\phi_{es}=1.77^{+0.24}_{-0.21}$(stat)~$^{+0.09}_{-0.10}$(syst)$\times10^6$~cm$^{-2}$s$^{-1}$.
These measurements are in agreement with previous solar neutrino flux
measurements, and with neutrino oscillation model results.  Including
these flux measurements in a global analysis of solar and reactor
neutrino results yields an improved precision on the solar neutrino
mixing angle of $\theta~=~34.4^{+1.3}_{-1.2}$ degrees, and
$\Delta m^{2}=7.59^{+0.19}_{-0.21} eV^2$.

\end{abstract}

\maketitle

\thispagestyle{fancy}


\section{THE SNO DETECTOR}

The Sudbury Neutrino Observatory (SNO) \cite{snonim} is a low
background neutrino detector $\sim$2~km (6000 mwe overburden)
underground in the Vale Inco Creighton nickel mine in Sudbury, Canada.
The detector consists of 1000 tonnes of D$_2$O in a 12m diameter
acrylic vessel surrounded by an inner shield of 1700 tonnes of
H$_{2}$O.  At the edge of the inner light water, a support structure
holding about 9500 photomultiplier tubes (PMTs) provides 54\%
coverage.  An additional outer shield of 5300~tonnes of light water
surrounds the PMTs.

The SNO detector detects the neutrino reactions:
$\nu_x+e^-~\rightarrow~\nu_x+e^-$ (ES), $\nu_e+d~\rightarrow~p+p+e^-$
(CC), and $\nu_x+d~\rightarrow~p+n+\nu_x$ (NC).  The SNO detector
provides the unique detection of the neutrons in the NC reaction by
three different methods, one for each phase of SNO running.  The first
phase detected gamma rays from the triton production in the detector
($n+d~\rightarrow~t+\gamma+6.25~MeV$)\cite{snod2o}.  In SNO's second
phase NaCl was added to the heavy water, and increased the neutron
capture through neutron capture on the Cl
($n+^{35}C~\rightarrow~^{36}Cl+\gamma+8.6~MeV$), where the 8.6 $MeV$ is
the sum of a cascade of gamma rays\cite{snosalt}.


For the final phase of SNO reported on here the neutron is detected
when it is captured in an array of 36 $^3$He proportional
counters (NCDs) via the reaction
$n+^3He~\rightarrow~p+t+0.76~MeV$\cite{ncdnim}\cite{ncdprl}.  The NCD
phase measurement separates the NC and CC signal detection which
significantly reduces the CC spectrum contamination by the 6.25 $MeV$
neutron captures on deuterium.  The NCD phase is more complex however
since $\sim$10\% of the Cherenkov light is blocked by the array, the
radioactivity of the counters adds a non-negligible background, and
the signal rate of $\sim$1000 neutrons/year is fairly low.

\section{NCD DETECTOR ENERGY SPECTRUM CALIBRATION}

The neutron energy spectrum is measured with calibration data from a
$^{24}$NaCl brine that produces neutrons by the gamma capture on
deuterium ($\gamma+d~\rightarrow~p+n$).  The spectrum is characterized
by a peak at 0.76 $MeV$ with features at 0.57 $MeV$ and 0.19 $MeV$
where either only the proton or triton are seen in the proportional
counter.

The $^{24}$NaCl brine is the calibration source most like the neutrons
produced from solar neutrinos since it can be uniformly distributed in
the D$_2$O, and it also provides a measurement of the neutron
detection efficiency ($0.211\pm0.007$).  The mixing of the brine can
be seen by looking at the light output from different parts of the
detector, and only data from after the brine was uniformly mixed was
used.  In addition, the detection efficiency from the MCNP \cite{MCNP}
Monte Carlo code yielded an efficiency of 0.210(3). Finally a
time-series based analysis using neutron bursts from a $^{252}$Cf
source confirmed these neutron efficiency measurements.

A simulation of the NCD detector was used to model the energy spectrum
from background alphas from U, Th, and Po in the nickel of the NCD
walls.  The model included effects of the energy loss, multiple
scattering, electron-ion pair generation, electron drift and
diffusion, electron multiple scattering, ion mobility, electron
avalanche, space charge, signal generation, and a detailed propagation
through the electronics.  The Monte Carlo simulation was tuned for the
surface to bulk alpha ratio, energy scale, energy resolution, alpha
depth, and contributions from different parts of the NCD using the
alphas above 2 $MeV$.  The simulation was found to reproduce the pulse
width and energy spectrum very well, including the effects of alphas
from the NCD anode wires.

Instrumental events in the NCD detector were easily separated from
ionization events using an amplitude versus energy cut.  Six of the 36
NCD strings with high instrumental rates were removed from the
analysis.  Two probability distribution functions for the instrumental
backgrounds were included in the signal-extraction to fit for an
unconstrained number of instrumental events.

\section{BLIND ANALYSIS}

Three blindfolds were implemented on the NCD phase measurement.  One
month of the data was open for analysts to tune cuts.  A hidden
fraction of neutrons that follow muons were added to the data, and an
unknown fraction of candidate events were omitted.  Detailed internal
documentation was reviewed by topic committees before the box was
opened to reveal the true solar neutrino flux measurement.  

The box was opened on May 2, 2008, and the results are presented as
found after correcting two inconsistencies.  The three separate
signal-extraction codes had to correct pilot errors on the inputs to
the final fit, which resulted in no change in the central values, and
made the final uncertainties reported agree.  An incorrect algorithm
in fitting the peak value of the ES posterior distribution was
replaced.

\section{NEUTRON BACKGROUNDS}

The neutron backgrounds were measured for the D$_2$O radioactivity,
atmospheric neutrinos, $^{16}$N neutrons, NCD counter neutrons from
the bulk of the counters, from hot-spots on the counters, and from the
NCD cables.  In addition, backgrounds from the acrylic-vessel, reactor
neutrinos, and other sources were included in the background
estimates.  The neutron backgrounds are summarized in the following
Table, and were included in the signal extraction broadening the
uncertainties in the measured solar neutrino fluxes. 

\begin{table}[ht]
\caption{Table of neutron backgrounds in the PMT and NCD data.}
\centering
\begin{tabular}{ l | c | c }
\hline\hline
Source & PMT neutrons & NCD neutrons  \\
\hline
\hline
D$_2$O radioactivity & 7.6$\pm$1.2 & 28.7$\pm$4.7 \\
Atmospheric $\nu$, $^{16}$N & 24.7$\pm$4.6 & 13.6$\pm$2.7 \\
Other backgrounds & 0.7$\pm$0.1 & 2.3$\pm$0.3 \\
NCD bulk PD, $^{17,18}$O($\alpha$,n) & 4.6$^{+2.1}_{-1.6}$ & 27.6$^{+12.9}_{-10.3}$ \\
NCD hot-spots & 17.7$\pm$1.8 & 64.4$\pm$6.4 \\
NCD cables & 1.1$\pm$1.0 & 8.0$\pm$5.2 \\
External-source neutrons & 20.6$\pm$10.4 & 40.9$\pm$20.6 \\
\hline
Total & 77$^{+12}_{-10}$ & 185$^{+25}_{-22}$ \\
\hline
\end{tabular}
\end{table}

\section{SIGNAL EXTRACTION METHODS}

Parameter estimation, and estimation of the uncertainties on all fit
parameters (both fluxes and nuisance parameters for systematics) is
done with a Metropolis algorithm Markov-Chain Monte Carlo
(MCMC)\cite{mcmc}.

In SNO's previous signal extractions, the negative log-likelihood
(NLL) function was simply minimized with respect to all parameters to
get the best-fit value, and the curvature of -$\log(L)$ at the minimum
was used to determine the uncertainties.  The floating systematics
approach also uses a minimization, although with additional nuisance
parameters added to account for systematic uncertainties.

Minimizing the NLL is very challenging for the 27 flux parameters
($\phi_{nc}$,$\phi_{cc1...13}$,$\phi_{es1...13}$) and 35 systematic
parameters in the fit.  The systematic parameters include PMT
reconstruction uncertainties estimated from calibration data
\cite{pmtcalib}, both PMT and NCD efficiencies, NCD Monte Carlo and
NCD instrumental uncertainties. In addition because the likelihood
function can be a bit choppy near the minimum, traditional minimizers
such as MINUIT run into trouble and often will not converge in
reasonable periods of time.

The MCMC method gets around this problem by interpreting NLL as the
negative log of a joint probability distribution for all of the free
parameters.  We then integrate over all nuisance parameters to
determine the distributions for the fluxes.  The origins of this
procedure go back to Bayesian probability theory, and in fact our
approach could be considered to be a Bayesian analysis with uniform
priors assumed for the fluxes.

The advantages of the MCMC method are twofold.  First, it converges
much faster than a 50+ parameter MINUIT minimization.  Rather than
minimizing over parameters, we in fact integrate over nuisance
parameters. Second, since we integrate over nuisance parameters with
the MCMC instead of trying to find a best-fit point, we are
insensitive to and in fact average over choppiness in the NLL that
would interfere with finding a minimum.  Both the speed of convergence
and the insensitivity to numerical noise in the NLL means that the
MCMC method is better suited to handling large numbers of nuisance
parameters.

\section{RESULTS}

The final corrected solar neutrino fluxes above a 6 $MeV$ Kinetic
Energy threshold from the unblinded NCD data are:
$\phi_{nc}~=~5.54^{+0.48}_{-0.46}\times10^6$~cm$^{-2}$s$^{-1}$,
$\phi_{cc}~=~1.67^{+0.08}_{-0.09}\times10^6$~cm$^{-2}$s$^{-1}$, and
$\phi_{es}~=~1.77^{+0.26}_{-0.23}\times10^6$~cm$^{-2}$s$^{-1}$.  The
correlation between $\phi_{cc}$ and $\phi_{nc}$ was only -0.19 in the
NCD phase fit including all systematic uncertainties.  The $\phi_{es}$
is a 2.2 sigma lower than the Super Kamiokande measurement, but the
full set of fluexes has a probability of 32.8\% of being consistent
with the six other flux measurements from all of the SNO phases.

\begin{figure}[htb]
\includegraphics[width=85mm]{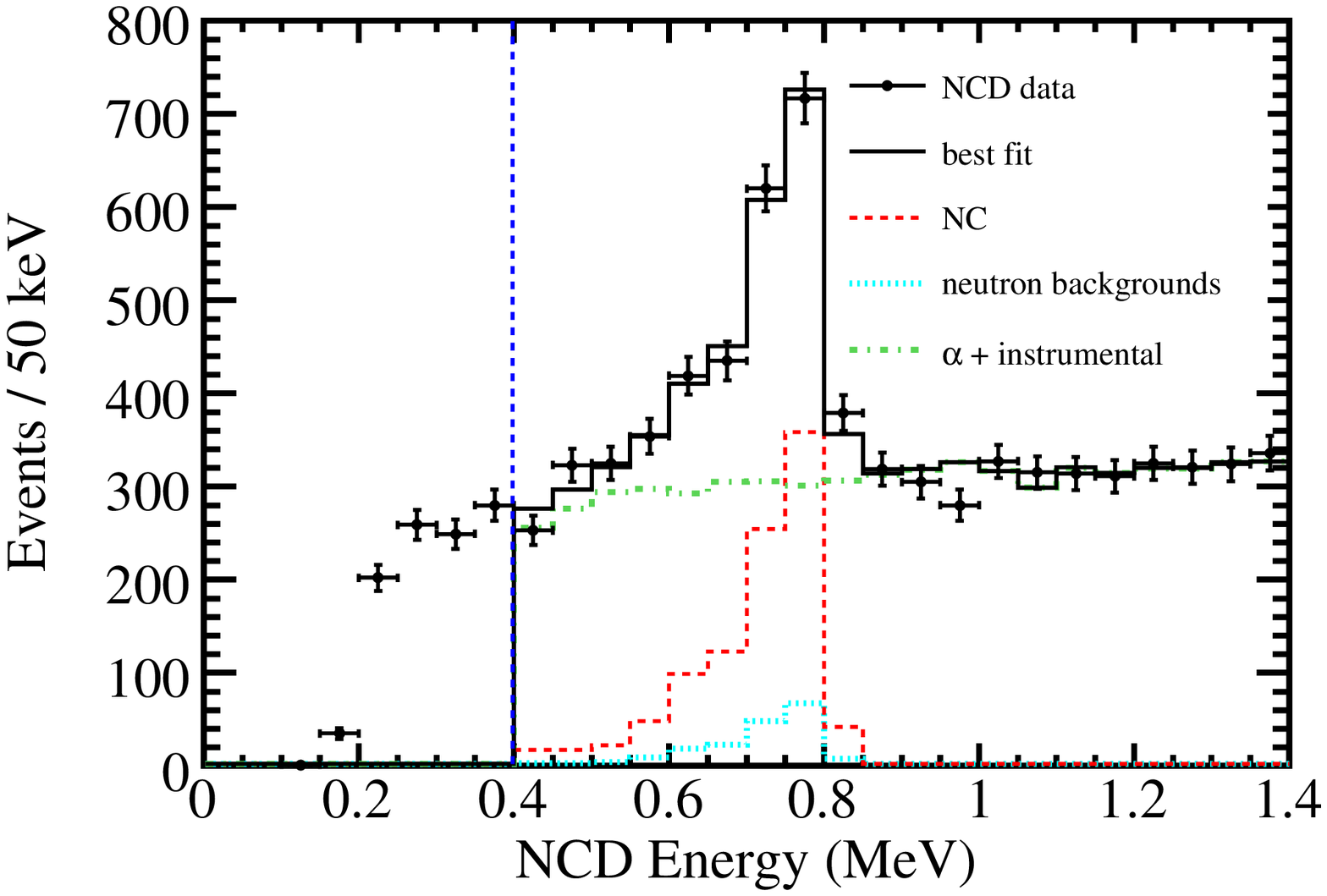}
\includegraphics[width=85mm]{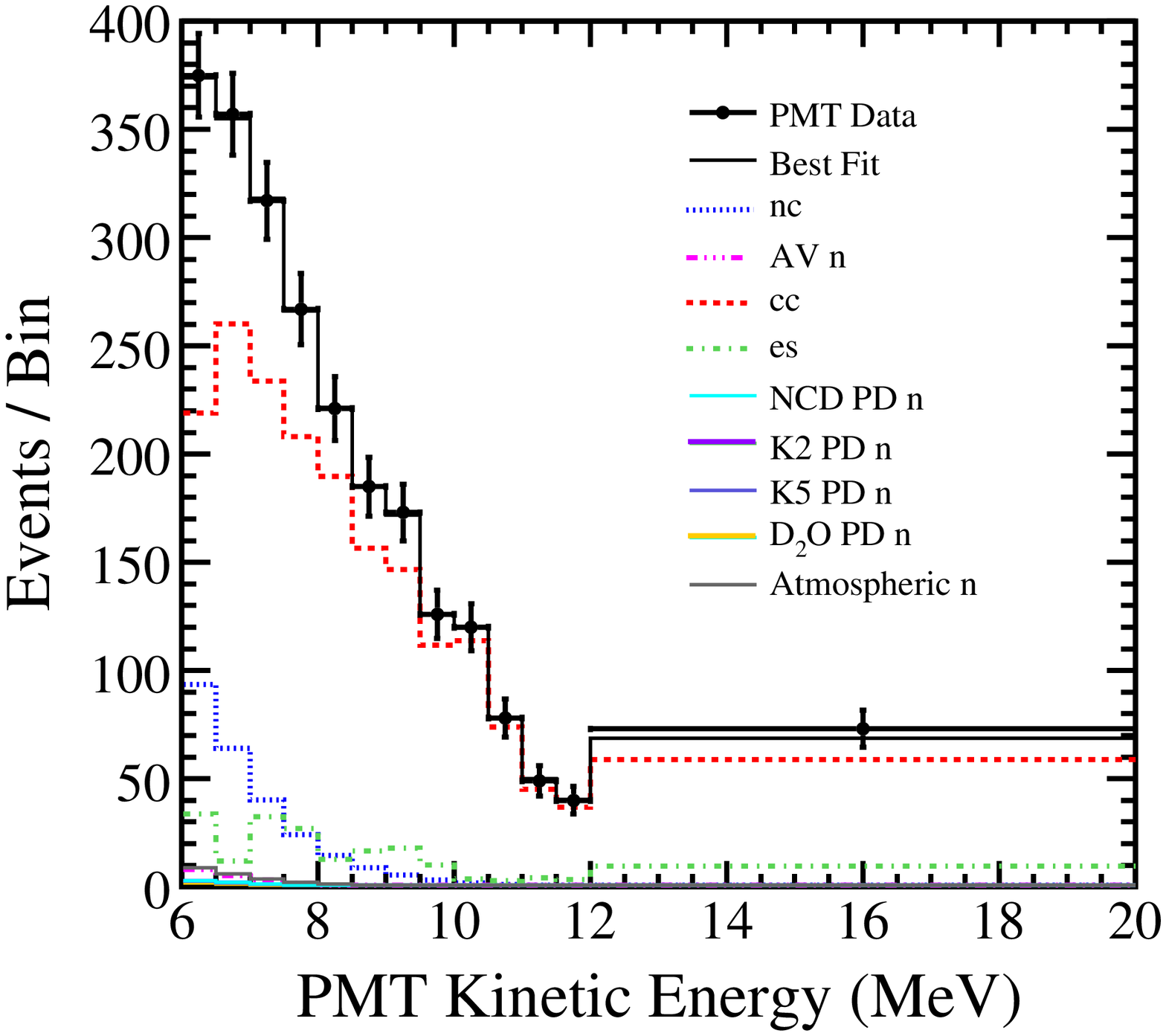}
\includegraphics[width=85mm]{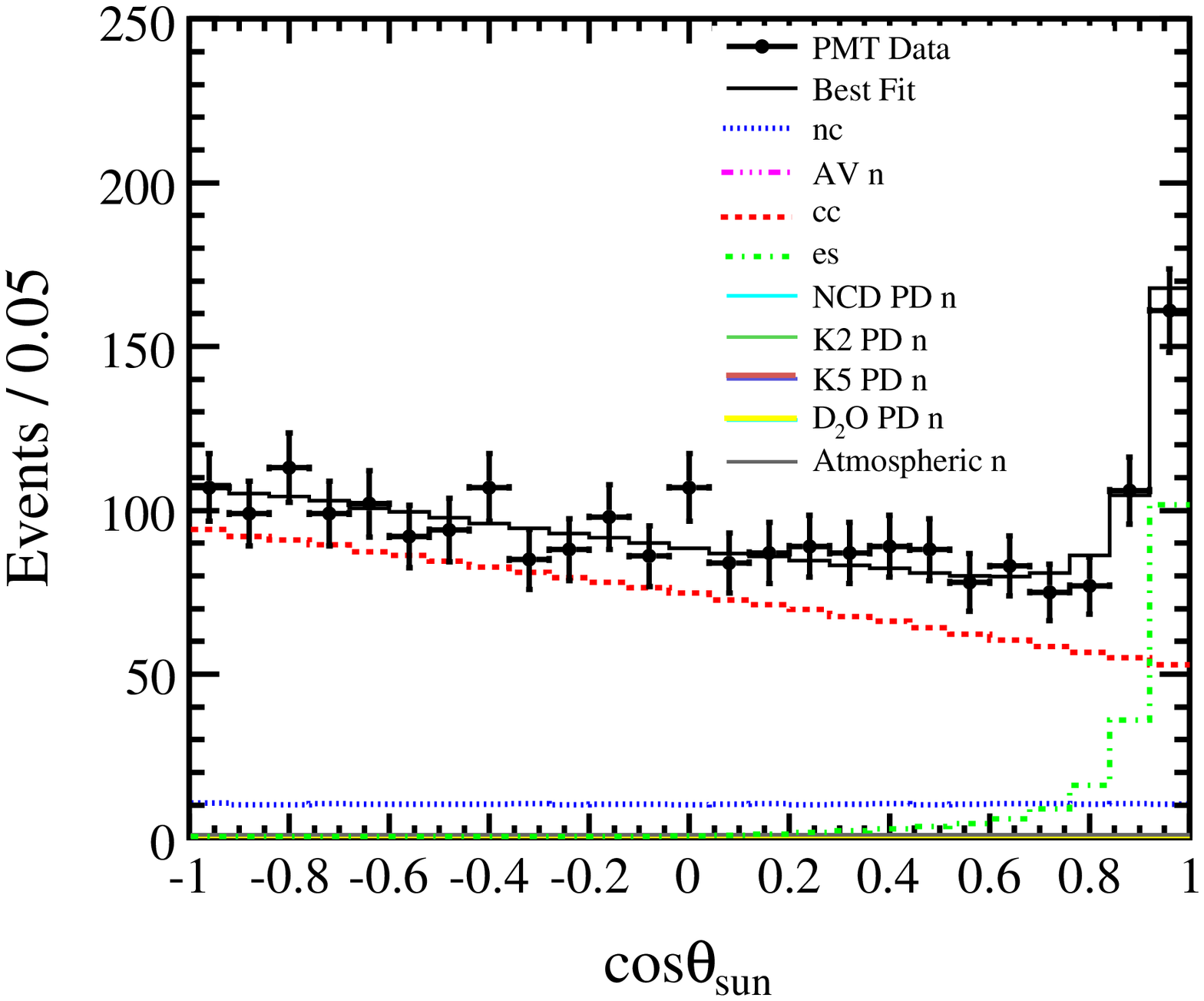}
\includegraphics[width=85mm]{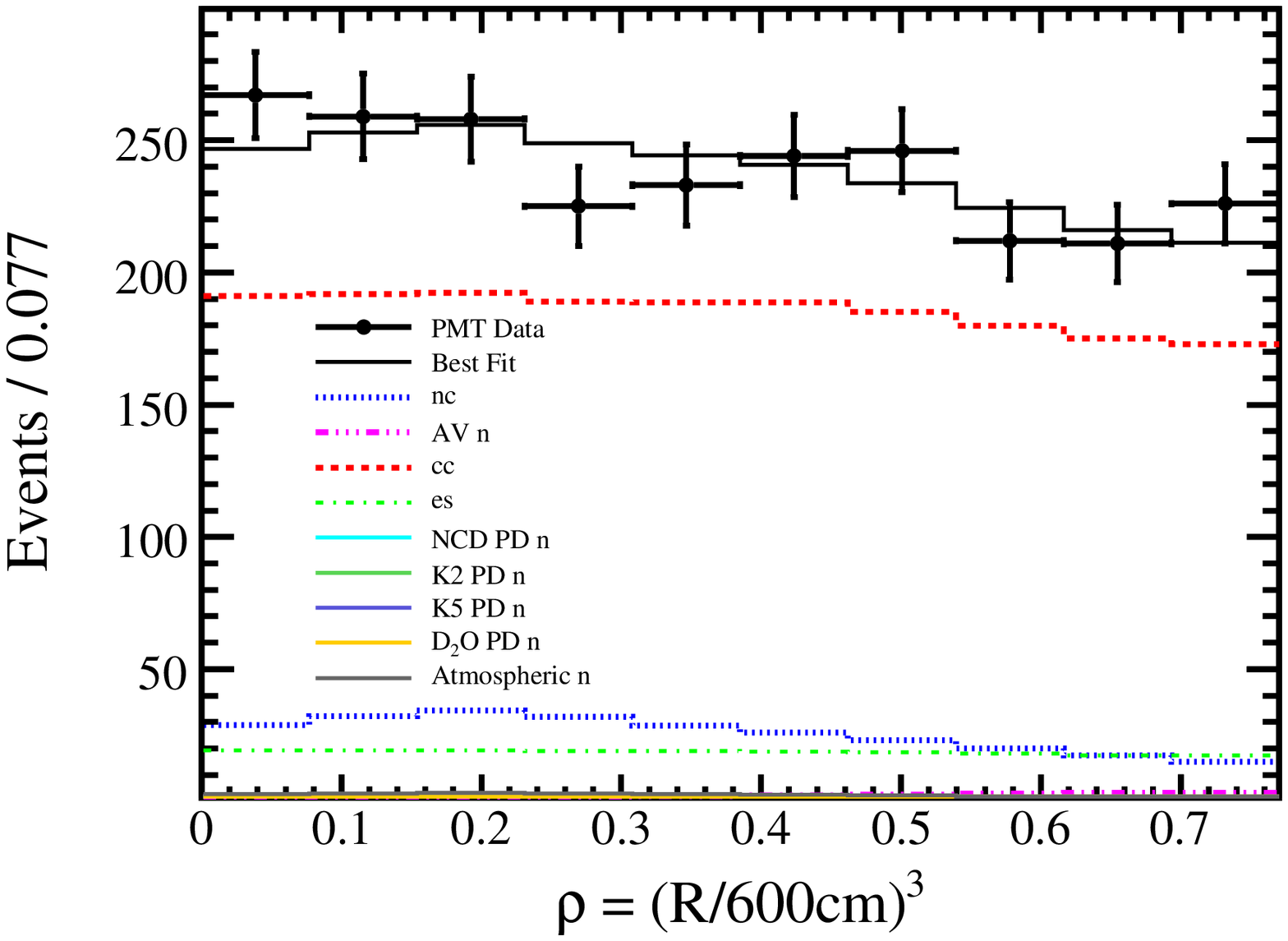}
\caption{\label{bestfitfigure}These figures show the NCD energy
  spectrum fit (top left), the PMT energy unconstrained fit (top
  right), the PMT angle to the sun fit (bottom left), and the PMT
  radial position fit (bottom right).}
\end{figure}

The NCD energy fit, unconstrained PMT energy fit, PMT angle to the
Sun, and radial position fits are shown in FIG.~\ref{bestfitfigure}.
It can be seen in the PMT energy fit that the number of neutrons in
the PMT is considerably less than either the SNO D$_{2}$O or salt
phases, and provides the better CC separation than in those phases.

Including the NCD phase flux measurements in a global oscillation
analysis results in an improved precision on the solar neutrino mixing
angle of $\theta~=~34.4^{+1.3}_{-1.2}$ degrees, and
$\Delta~m^{2}=7.59^{+0.19}_{-0.21} eV^2$.

\begin{acknowledgments}
SNO gratefully acknowledges NSERC, Industry Canada, NRC, Northern
Ontario Heritage Fund, Vale INCO, Atomic Energy of Canada, Ontario
Power Generation, High Performance Computing Virtual Laboratory,
Canada Foundation for Innovation, Canada Research Chairs, Westgrid, US
Department of Energy, NERSC PDSF, UK STFC, and Portugal FCT.
\end{acknowledgments}

\end{document}